# Beam smoothing based on prism pair for multistep pulse compressor in PW lasers

SHUMAN DU[1, 2], XIONG SHEN[1], WENHAI LIANG[1, 2], PENG WANG[1], JUN LIU[1, 2, *]

[1] *State Key Laboratory of High Field Laser Physics and CAS Center for Excellence in Ultra-intense Laser Science, Shanghai Institute of Optics and Fine Mechanics, Chinese Academy of Sciences, Shanghai 201800, China*

[2] *University Center of Materials Science and Optoelectronics Engineering, University of Chinese Academy of Sciences, Beijing 100049, China*

\* *jliu@siom.ac.cn*

**Abstract:** Ultra-short ultra-intense laser provides unprecedented experimental tools and extreme physical conditions to explore frontier secrets of nature. Recently, multistep pulse compressor (MPC) was proposed to break through the limitation of the size and damage threshold of the grating in the compressor during the realization of higher peak power laser. In the MPC methods, beam smoothing in the pre-compressor is a very important process. Here, beam smoothing based on prism pair were studied technically, in which both the spatial profiles and the spectral dispersive properties were analyzed in detail. The simulation results show clearly that the prism pair can effectively smooth the laser beam. Furthermore, the beam smoothing is much more efficiency with shorter separated distance if two prism pairs are arranged to induce spatial dispersion at one direction or two directions. The results of beam smoothing here will help the optimized optical designs in all PW laser systems to improve their output and running safety.

## 1. INTRODUCTION

Ultra-intense femtosecond laser pulses provide unprecedented extreme conditions for the exploring of frontier secrets of nature, which help to reveal new phenomena of matters in the strong-field ultrafast laser field and have attracted many attentions [1-3]. Since the chirped pulse amplification (CPA) method was put forward in 1985 [4], more than 50 lasers with peak power reached hundreds terawatt (TW) or even petawatt (PW) in the world [5]. Several hundred level PW femtosecond (PW-fs) laser systems are in planning or building in Europe, Russian, USA and China, they are ELI-200PW [6], XCLES-200PW [7], OPAL-75PW [8] and SEL-100PW [9] facilities, respectively. In a PW-fs laser system, the laser pulse is firstly stretched from femtosecond to nanosecond and amplified by using the CPA or the optical parametric chirped pulse amplification (OPCPA) [10] techniques, where the chirped nanosecond pulse is compressed to ultra-short femtosecond pulse by using a grating-based pulse compressor in the final stage.

In a PW-fs laser system, the compression system is usually a Treacy four-grating compressor [11], where the first grating surface bears the largest input pulse energy, and the fourth grating bears the shortest pulse duration or the highest peak power, so the damage threshold of the first and fourth gratings limit the maximum input and output pulse energy, respectively. The output laser beam of the main amplifier usually owns large laser spatial intensity modulation (LSIM) due to the pump laser and hot spots appear in the beam. The reason of appearance of hot spots is the diffuse reflection of dust or the defects of optical components. These strong LSIM or hot spots will reach the damage threshold of the grating ahead of the average intensity and limit the output pulse energy at the end of the laser system.

Recently, a novel optical design named multistep pulse compressor (MPC) was proposed to increase the input/output pulse energy of a grating-based pulse compressor [12]. In the MPC method, beam smoothing based on prism pair is a very important process used as the pre-compressor step. However, this beam smoothing process has not been studied in detail. In this paper, the beam smoothing process is studied technically by using prism pair induced spatial dispersion to the laser beam. Since the prism pair will also introduce spectral dispersion to the laser pulse with relative broadband spectrum, not only the spatial properties but also the spectral dispersion is analyzed in the paper. The simulation results demonstrated the beam smoothing based on prism pairs can be used to reduce the laser spatial intensity modulation ratio (LSIMR) of the laser beam on the surface of the first and last grating in the compressor. Thus, the pulse energy of the incident and output laser beam can be increased without damaging the grating. Furthermore, this beam smoothing based prism pair can also be extended used before the main amplifier to smooth the incident laser beam before the big crystal for high energy amplification to protect the expensive crystal.

## 2. PRINCIPLE OF PRISM PAIR INDUCING SPATIAL AND SPECTRAL DISPERSIONS

Prism pair is a set of commonly used dispersion element, which has the advantages of simple composition, flexible operating, low loss, and precisely adjustable spectral dispersion. Negative spectral dispersion is usually induced by prism pair to compensate the positive dispersion so as to achieve ultrashort laser pulse with small laser beam. Usually, only the induced negative spectral dispersion was useful and the induced spatial dispersion was harmful in many these kinds of applications, where the laser beam will pass through the prism pair two times, forward and backward, to compensate the prism pair induced spatial dispersion automatically.

Recently, it was found that the induced spatial dispersion was also very useful in laser microscopy and laser micro-machining applications [13, 14]. In these applications, the laser beam will be introduced amount of spatial dispersion intentionally by using a grating/prism pair, and then focus it by using lens or other focusing optics. The purpose of this special optical design is to achieve the highest focal intensity at the focal point, in which the most important aspect is to ensure the laser intensity near the focal point will be decreased rapidly. In this way, the optical section capability will be improved near the focal point which will increase the resolution and contrast of microscopy or micro-machining. This method was named spatiotemporal focusing [15].

In very recently, we found this spatiotemporal focusing can also be used in ultrahigh peak power laser system with large beam size. In this MPC optical design, new application was discovered with the induced spatial dispersion by using the grating or prism pair. This time, the induced spatial dispersion was used to smoothing the laser beam to reduce its LSIMR [12].

In spatial domain, a laser beam pass through a prism will introduce angular dispersion which will extending the laser beam in the angular dispersion direction owing to broad spectrum, while another prism with the same optical parameters was located parallel to the first one to collimate the output laser beam. The optical setup of prism pair with two right angle prisms for laser beam smoothing is shown in Fig. 1(a), where $L$ is the perpendicular distance between the two prisms, α is the apex angle of the two prisms, $\theta_s$ and $\theta_l$ are the exit angles of the shortest wavelength and the longest wavelength, respectively. $d$ is the induced spatial dispersion width, w0 and w are the input and output laser beam widths along the X axis, respectively. The induced spatial dispersion width d can be expressed simply as $d = L(tan\theta_s - tan\theta_l)cos\alpha$. Considering the relationship between the exit angle of the beam $\theta$ and the apex angle of the prism as well as that between $\theta$ and the refractive index of prism pair at different wavelengths, the induced spatial dispersion width $d$ can also be expressed as:

$$d = L(\frac{M_s}{\sqrt{1-M_s^2}} - \frac{M_l}{\sqrt{1-M_l^2}})cos\alpha \qquad (1)$$

$$M_s = n_s sin(\alpha - \arcsin\frac{sin\delta}{n_s}) \qquad (2)$$

$$M_l = n_l sin(\alpha - \arcsin\frac{sin\delta}{n_l}) \qquad (3)$$

Where $\delta$ is the incident angle of laser beam, $n_s$ and $n_l$ are the refractive index of prism pair at the shortest wavelength and the longest wavelength, respectively.

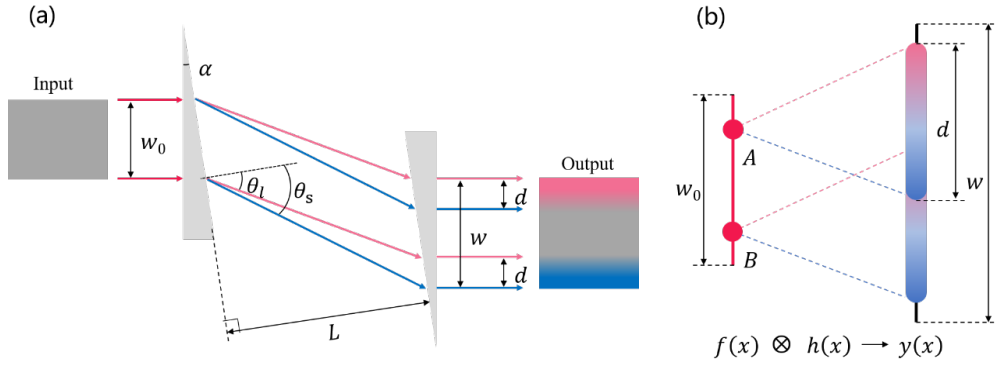

Fig. 1 (a) The optical setup of prism pair with two right-angle prisms for laser beam smoothing. $L$: the perpendicular distance between the two prisms, $\alpha$: the apex angle, $d$: the induced spatial dispersion width, $w_0$ and $w$ are the input and output laser beam widths along the X axis, respectively. $\theta_s$ and $\theta_l$: the exit angles of the longest wavelength and the shortest wavelength, respectively. (b) Principle of beam smoothing due to the convolution between spatial and spectral distribution of the laser beam. A and B are two hot spots with full spectra in the laser beam.

From the above expresses, it can be seen that $d$ are related to the distance $L$ between the two prisms, the apex angle of prism $\alpha$ and the exit angles of beam $\theta$. The refractive index of prism pair is constant at same wavelength, thus changing the range of spectrum has a great influence on $d$.

Note that the center part of the output laser beam with full bandwidth spectrum and no spatial dispersion for a laser beam with big size, while the laser beam on both sides own spatial dispersion.

Due to the angular dispersion of the prism pair, the light with different wavelength components will be output with different angles from the first prism, and distributed at different position after being collimated by the second prism. Different spots nearby in the incident laser beam will overlapped their spatio-spectral distribution in the output beam, therefor the intensity of the output laser beam will be redistributed. Figure 1(b) demonstrate the principle of this redistribution by using two nearby input light spots A and B. Assume the laser spectrum are the same for all spots in the incident laser beam, both A and B will be extended from a spot to a line with a length of d if the two prisms separated by $L$ distance. If the distance between A and B smaller than $d$, there are spatial overlapping with different wavelengths in the output laser beam for A and B. Obviously, the process shown in Fig.1(b) which inducing spatial dispersion by angular dispersion can also be explained by using convolution, as shown in the following equation.

$$y(x) = \sum_{-\infty}^{+\infty} f(n)h(x-n) = f(x) \otimes h(x) \tag{4}$$

Where $f(x)$ is the incident laser intensity distribution function along the X axis, $h(x)$ function is the spectral intensity profile projecting on the X axis, $y(x)$ is the obtained output smoothed beam profile on the X axis. As we know that the convolution operator owns the ability of smoothing figures or signals. The final laser beam intensity is obtained by accumulating the intensity of light that arriving at the same spatial position through the prism pair. As a result, the output laser beam is smoothed owing to the convolution and intensity redistribution.

How much spatial intensity modulation ratio improvement can be obtained by using this beam smoothing process? Since hot spots is the typical damage risk for high peak power laser which owing small beam size and local high intensity. We assume points A and B own a diameter of $2 \times r$, as a result, the peak power of the two hot spots are decreased by about $2 \times r \times d/(\pi \times r^2) = 2/\pi \times d/r$ times. After summing up all the corresponding spatio-spectral intensity values at every point on the extended output beam line, the beam smoothing effect of the output laser beam is related to the $d/r$ ratio. It means that increasing $d$ or reducing $r$ have a better effect on smoothing the output laser beam. In other words, inducing a larger spatial dispersion width d to hot spots with high spatial frequency 1/$r$ is easier to be smoothed.

Except the spatial dispersion, single pass of prism pair will also induce spectral dispersion to the input laser pulse. For a PW laser, the laser pulse is usually chirped to nanosecond to avoid laser induced damage in the amplification crystal. In comparison to the spectral dispersion of the input laser beam, the induced spectral dispersion is very small. As for tens femtosecond transform limited laser pulse with broadband spectrum, the induced small spectral dispersion is also very important for the well compression of the laser pulse. According to many previous works, the induced spectral dispersion by single pass of prism pair can be expressed as [16]:

$$GDD = \frac{\lambda^3}{2\pi c^2} \frac{d^2 p}{d\lambda^2} \tag{5}$$

$$TOD = -\frac{\lambda^4}{4\pi^2 c^3}\left\{3\frac{d^2 p}{d\lambda^2} + \lambda\frac{d^3 p}{d\lambda^3}\right\} \tag{6}$$

Where GDD and TOD are the second-order dispersion and the third-order dispersion, respectively. $\frac{d^2 p}{d\lambda^2}$ and $\frac{d^3 p}{d\lambda^3}$ are the second-order and the third-order derivatives of optical path to wavelength. $\lambda$ is wavelength and $c$ is speed of light. Moreover, it was found that the combination of prism pair and grating pair can help to compensate the third-order dispersion, and then help to achieve shorter compressed pulse duration because the prism pair and grating pair owing same symbol GDD and opposite symbol TOD[11, 16]. Note that the prim is much bigger with longer material dispersion for big beam size, then the spectral dispersion will be different and affect by the material dispersion in comparison to the prism pair used for small beam size, which will be discussed in the following section.

## 3. SIMULATION AND RESULTS IN SPATIAL DOMAIN

In the specific simulation, the spatial profile of the incident beam is set to be a 10th order flat top super-

Gaussian beam, the beam size is 370×370 mm, and the LSIMR of the incident laser beam is set to be about 2.0. The spectral shape of the input beam is set to 7th order flat top super-Gaussian and the spectral range is 825 nm-1025 nm. The incident beam is sampled at an interval of 1 mm in space.

Using ray tracing method, the spatial distribution of different wavelength components in the sampling point with full spectrum is calculated one by one. Finally, the output laser beam with new spatial distribution is obtained. By changing the parameters of prisms and various combinations of prism pairs, the smoothing effect of output laser beam will be discussed in detail.

2.1 One prism pair

In this paper, the beam smoothing effect of right-angle prisms with apex angles of 15°, 20°, 25°and 30° is simulated, where the value of LSIMR is used to characterize the effect of beam smoothing. When the laser beam is incident at zero degree on the right-angle prism pairs with different apex angles, the variations of LSIMR and the induced spatial dispersion width as the distance of the prism pair changes are shown in Fig. 2(a) and Fig. 2(b), respectively.

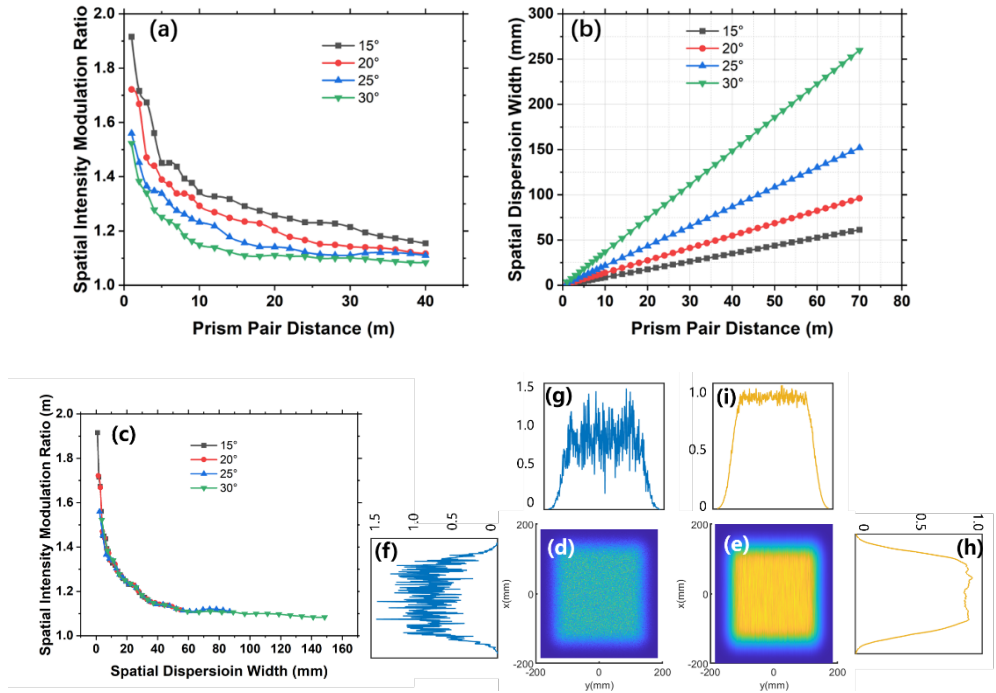

Fig.2. (a)The variations of LSIMR of the laser beam, (b) the variations of the induced spatial dispersion width, (c)the variations of LSIMR with the induced spatial dispersion width (d) The two-dimensional spatial intensity profiles of the input laser beam and (e) the smoothed output laser beam. The one-dimensional spatial intensity profiles (f-g) of the input laser beam and (h-i) the smoothed output laser beam.

In Figure 2 (a), LSIMR decreases rapidly from 2.0 to about 1.3, indicating that the prism pair by inducing spatial dispersion can effectively smooth the beam. However, LISMR drops from 1.3 to around 1.1 requires a long distance of right-angle prism pair. By comparing the curves of LISMR of different prism apex angles, it is concluded that achieving the same LISMR requires prism pair distance of apex angle of 30° is much shorter than that of 15°. In Fig.2 (b), the induced spatial dispersion width d increases linearly with the perpendicular distance of prism pair $L$. Since the exit angle of the light passing through the prism pair is related to the apex angle of the prism, the same induced $d$ requires the distance of prism pair $L$ for apex angle of 30° is much shorter than 15°. For prisms with different apex angles, the curves of LISMR show the same downward trend, and the trend changes are almost the same, indicating that LISMR is closely related to the induced *d* and the results is in Fig. 2 (c), which further confirms that the prism pair can effectively smooth the laser beam.

By comprehensively comparing the induced spatial dispersion width and LISMR for prism pairs with different apex angles, as long as the induced spatial dispersion width d reaches about 60 mm, LISMR is reduced to about 1.1 and the laser beam will be well smoothed. Fig. 2 (d-i) show the one-dimensional and the two-dimensional spatial intensity profiles of the input laser beam and those of the smoothed output laser beam. The input laser beam is well smoothed on the X and Y axes and LISMR is about 1.1.

The above discussion is an ideal case with smoothing spectrum. However, there is no completely smoothing laser spectrum in the actual situation. Here, we simulate the laser beam smoothing with modulated spectra. Fig.3(a) shows the one-dimensional shape of ideal spectrum and modulated spectrum. The curve of LSIMR that the laser beam passing through a right-angle prism pair of apex angle of 20° with the increase of the distance of prism pair, as shown in Fig.3(b). According to the simulation results, the spectrum with intensity modulation does not have much influence on LSIMR. The reason is that the change of LSIMR is the convolution of the intensity of the incident laser beam and the intensity of the spectrum of laser beam. Fig. 3 (c-h) show the one-dimensional and the two-dimensional spatial intensity profiles of the smoothed output laser beam by the input laser beam with ideal spectrum and modulated spectrum passing through a distance of 8 meters prism pair of apex angle of 20°. From Figure 3(c-h), it can be seen more intuitively that the shape of spectrum has no effect on smoothing beam.

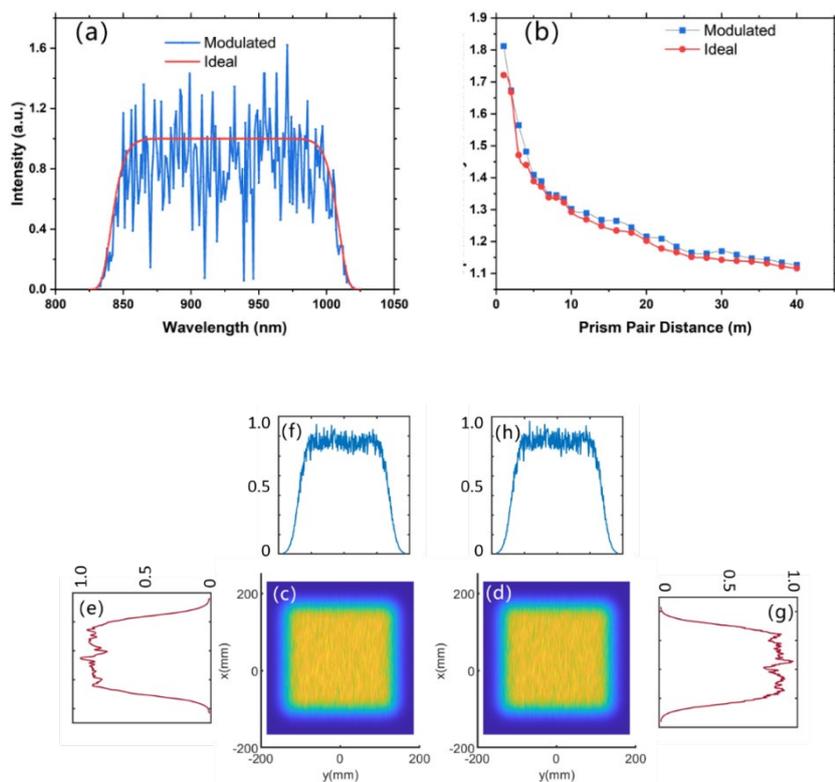

Fig.3(a) Super Gaussian spectrum and the modulated spectral, (b) The variation curves of the spatial intensity modulation ratio of the laser beam with the modulated spectrum and ideal spectrum, (c) The two-dimensional spatial intensity profiles of the output laser beam with ideal spectrum and (d) the modulated

spectrum. The one-dimensional spatial intensity profiles (e-f) of the output laser beam with ideal spectrum and (g-h) the modulated spectrum.

From the above analysis, the larger the induced spatial dispersion width, the better LSIMR decreases. And the larger the apex angle of the right-angle prism can help to achieve the larger induced spatial dispersion width. Nevertheless, the right-angle prism pair with large apex angle may have total reflection when it is incident at zero degree, and the dispersion is induced into the right-angle prism pair only on the oblique side. Whether to replace the right-angle prism pair with an isosceles prism pair can introduce a larger amount of dispersion? The optical setup and simulation results are shown in Fig.4(a) and Fig.4(b), respectively, where $\beta$ is the apex angle of isosceles prism. Compared with the simulation results of right-angle prism pair, the induced spatial dispersion width of isosceles prism pair with parallel incidence can be achieved by longer prism pair distance, and isosceles prisms with parallel incidence do not have a good smoothing beam effect than right-angle prisms. According to the formulas (1-3), the induced spatial dispersion width $d$ is greatly related to the apex angle and incident angle of the prism. The larger apex angle and zero incident angle, the better beam smoothing effect. A full-spectrum hot spot parallel enters the isosceles prisms, the introduced dispersion width is smaller than the induced dispersion width of right-angle prisms with zero incident when the apex angle $\beta = \alpha$. As a result, the shape of the prism pair has no effect on spatial dispersion.

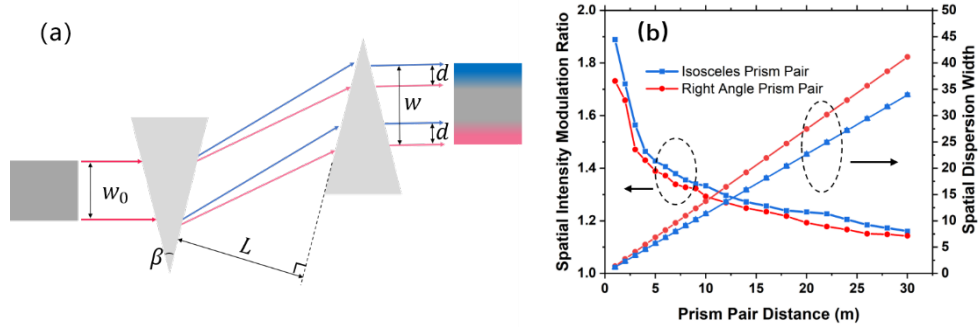

Fig.4(a) the optical setup of the isosceles prism pair for smoothing laser beam and (b) the variation curves of the spatial intensity modulation ratio and the variation curves of the induced spatial dispersion width.

2.2 Two prism pairs in one direction

From the simulation results of the laser beam smoothing by a single prism pair, LSIMR decreases to about 1.1 which requires the induced spatial dispersion width about 60 mm. It means that a long distance between prism pair is required to achieve these conditions. However, a long distance between prism pair will limit the application of experiments in reality. Here, an effective prism pair, where the visual single prism is combined with two prisms together to achieve a relative larger apex angle. Then, it is possible to increase the induced dispersion width with relative a small experiment space. The optical setup is shown in Fig.5 (a), where four prisms are exactly the same, $\varepsilon$ is the apex angle. $L1$, $L2$, $L3$ are the distances between prisms, and $L = L1 + L2 + L3$. Usually, $L1 \approx 0, L3 \approx 0$ and $L = L1 + L2 + L3 = L2$, which means the total distance between the first and the fourth prism. In this paper, $L1$ and $L2$ are both set to 0.05 meters, and the total distance $L$ is changed by changing $L2$. Note that there are many combinations of prisms and various $L1$, $L2$, $L3$. $L1$ and $L2$ are not necessarily equal, this is just the result of the following display. In a real experiment, the right-angle surface of the first prism is parallel to the right-angle surface of the second prism and the first prism is closed to the second prism. The distance between the first prism and the second prism can be adjusted by the translation stage. In the same way, the third prism and the second prism are placed in the same way as above. In order to keep the propagation direction of the input laser beam and output laser beam consistent, the exit surface of the second prism and the entrance surface of the third prism must be kept parallel. From simulations of various combinations of prisms, $L2$ contributes the most to spatial dispersion width $d$ and the contribution of $L1$ and $L2$ are the same. At this time, a single hot spot is extended into a line with a length of $d = d_{x1} + d_{x2} + d_{x3} \approx d_{x2}$ and a width of $2 \times r0$, the local intensity will be decreased by about $2(d_{x1} + d_{x2} + d_{x3})/(\pi r) \approx 2d_{x2}/\pi r$, where $d_{x1}$, $d_{x2}$, $d_{x3}$ are the induced spatial dispersion width along X axis by the prism distance of $L1$, $L2$ and $L3$ respectively.

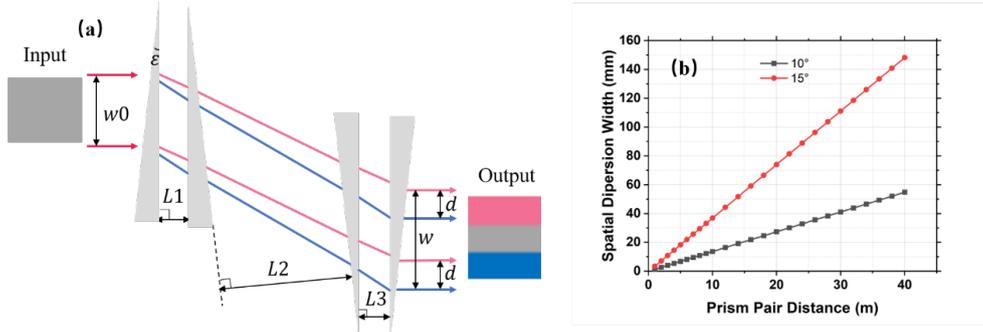

Fig.5 (a) The optical setup of two prism pairs in one direction, where ε is the apex angle. L1, L2, L3 are the distances between prisms. (b) the curves of the spatial dispersion width of the laser beam passing through two prism pairs.

The simulation results of spatial dispersion width d changed with $L$ are shown in Fig.5(b), where the apex angle of prisms are 10°and 15°. It is found that the spatial dispersion width of two prism pairs, which is equivalent to the spatial dispersion width of a single prism pair with twice the apex angle. It means that the two prism pairs with apex angle of 15°equivalent to a single prism pair with angle of 30°. For a small apex angle prism of 15°, when the spatial dispersion width reaches 60 mm or 10 mm, a single prism pair needs a distance of 68 m or 12 m, while two prism pairs only need a distance of 16 m or 3 m. It saves 52 m or 9 m of distance by comparing a single prism pair of 15°with induced 60 mm or 10 mm spatial dispersion width. Obviously, this combination of prisms not only saves spacing between prisms, but also retains the advantages of small apex angle prisms.

2.3 Two prism pairs in two directions

In the above simulation results, two prism pairs can effectively smooth the beam by inducing spatial dispersion to the X axis. In order to introduce angular dispersion in two axis and save experiment space, placing the other prism pair perpendicular to the first prism pair, which can induce spatial dispersion in the input beam along the Y axis direction. The optical setup is shown in Fig. 6(a). It need to note that, at this time, the single hot spot is extended into not a line but a area of $d_x \times d_y$ after passing through two prism pairs, where $d_x$ is the induced spatial dispersion width along X axis and $d_y$ is the induced spatial dispersion width along Y axis. As a result, the overall induced spatial dispersion area in the central spatial region with full spectral bandwidth is about $d_x \times d_y$. As for a single hot spot with a diameter of $2 \times r$, the local intensity will be decreased by about $d_x \times d_y/(\pi r^2)$. This ratio is only calculated for a single hot spot, which overall effect is the superposition of convolution integrals, and the lowest value is reduced to the average value, not infinitely reduced. Note that if the decreased intensity of the smoothed single hot spot is smaller than the average intensity, the beam will be smoothed to a LSIMR of about 1.0.

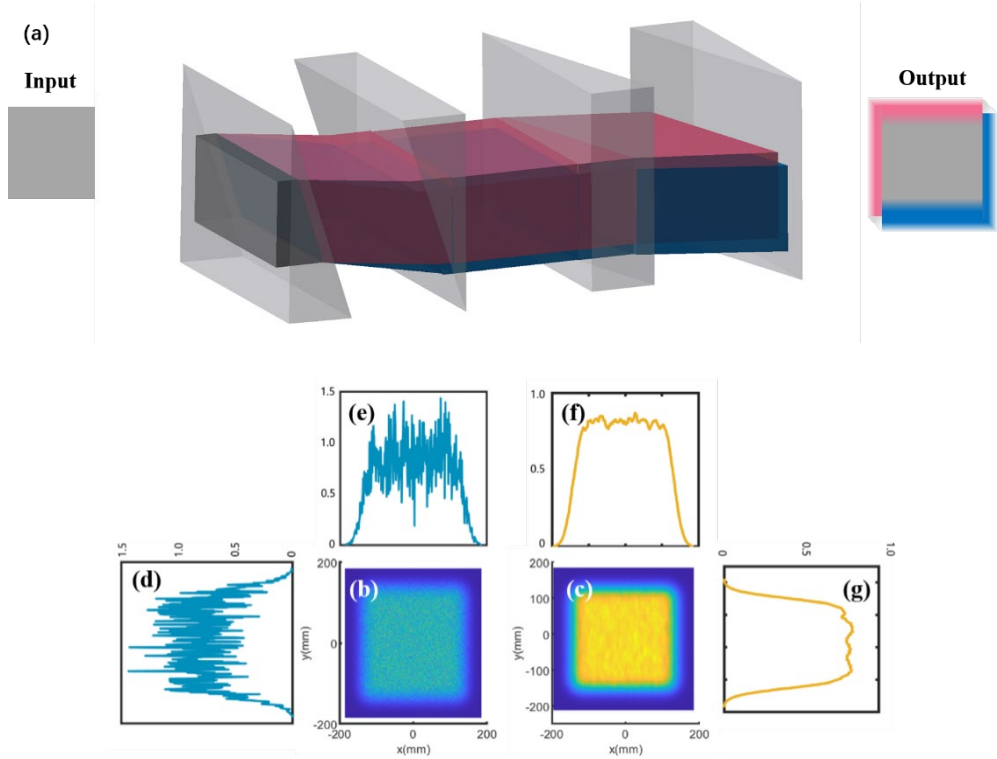

Fig.6(a) Optical setup with two prism pairs which are located in both X and Y axes. (b) the two-dimensional spatial intensity profiles of the input laser beam and (c) the smoothed output laser beam in X and Y axes. The one-dimensional spatial intensity profiles of (d-e) of the input laser beam and (f-g) the smoothed output laser beam in X and Y axes.

In order to maintain the advantage of a small apex angle of prism pair, a right-angle prism pair with a apex angle of 15° is chosen in the simulation inducing angular dispersion in both X and Y directions. From the above conclusions, the spatial dispersion width reaches 60 mm, the laser smoothing effect is great. Due to the introduce spatial dispersion width divide two parts in X and Y directions and the total spatial dispersion width $d = d_x \times d_y$ is constant, using the inequality $\sqrt{ab} \leq (a + b)/2$, it can be concluded that when $d_x = d_y$, the $d_x + d_y$ is the minimum. Where $a$ and $b$ are both positive real numbers. Thus, the distance between the first prism pair with apex angle of 15° is set to 9 m to introduce about 8 mm spatial dispersion width both along the X axis and the perpendicular distances between second prism pair is also 9 m. As a result, a total prism pair distance of about 18 m can smooth the laser beam to a LSIMR of 1.1 which will save the total optical space greatly. Fig.6 (b-g) shows the one dimensional and the two-dimensional spatial profiles of the input laser beam and those of the smoothed output laser beam in both X axis and Y axis, which the induced spatial dispersion width is 10mm along X axis and 6 mm along Y axis.

## 4. SIMULATION AND RESULTS IN SPECTRAL DISPERSION

The laser beam passing through the prism pair not only changes the spatial characteristics of laser beam, but also affect its temporal properties. The induced spectral dispersion by the prism pair of the output laser beam consists of two parts, one is the material dispersion of the prism pair itself, and the other is the negative dispersion caused by the angular dispersion of the prism pair. As for PW laser facility, the laser beam is relative large, and then the prism pair used for beam smoothing also own a big optical size which means a relative thick optical path in prisms. As a result, the induced positive dispersion by the material of prism pair is considerable which is different from the prism pair with small size. The positive dispersion caused by the material is as follows:

$$\varphi(\omega) = \frac{\omega}{c} n(\omega) l \tag{7}$$

$$GDD_m = \frac{d^2\varphi}{d\omega^2} = \frac{\lambda^3 l}{2\pi c^2} \frac{d^2 n}{d\lambda^2} \tag{8}$$

$$TOD_m = \frac{d^3\varphi}{d\omega^3} = -\frac{\lambda^4 l}{4\pi^2 c^3}\left\{3\frac{d^2 n}{d\lambda^2} + \lambda\frac{d^3 n}{d\lambda^3}\right\} \tag{9}$$

Where $GDD_m$ and $TOD_m$ are the positive dispersion of second-order and the third-order, respectively. $\frac{d^2 n}{d\lambda^2}$ and $\frac{d^3 n}{d\lambda^3}$ are the second-order and the third-order derivatives of refractive index to wavelength. $l$ is the distance traveled in the prism pair.

As for the introduced negative dispersion from the angular dispersion induced by the prism pair, the second-order dispersion GDD and third-order dispersion TOD are simulated based on the formulas (5) and (6), respectively. In the simulation, the apex angle of prism is 20°, the center wavelength is 925nm, and the beam size is 370×370 mm. The spectral shape of the input beam is set to 7th order flat top super-Gaussian and the spectral range is 825 nm-1025 nm. As we know, the GDD and the TOD is related to the prism insertion, the first prism insertion is 10 mm and the second prism insertion is 400 mm in this simulation. For a single prism pair, the induced GDD and TOD with center wavelength 925nm varying with the increases of the prism pair distance, as shown in Fig. 7(a) and Fig.7(b), respectively. Fig.7(b) is the enlarged part marked by the blue box in Fig.7(a). As we can see that the introduced GDD and TOD are both positive value at the beginning due the relative large prism and thick material induced positive dispersion. When the prism pair spaced by about 0.9 m, the angular dispersion induced negative dispersion and the material induced positive dispersion is balanced which induce a nearly zero GDD and TOD. In the case, the induced spatial dispersion width $d$ is about 1.3 mm. As the increasing of the prism pair distance, the GDD and TOD tuning to negative and increasing linearly.

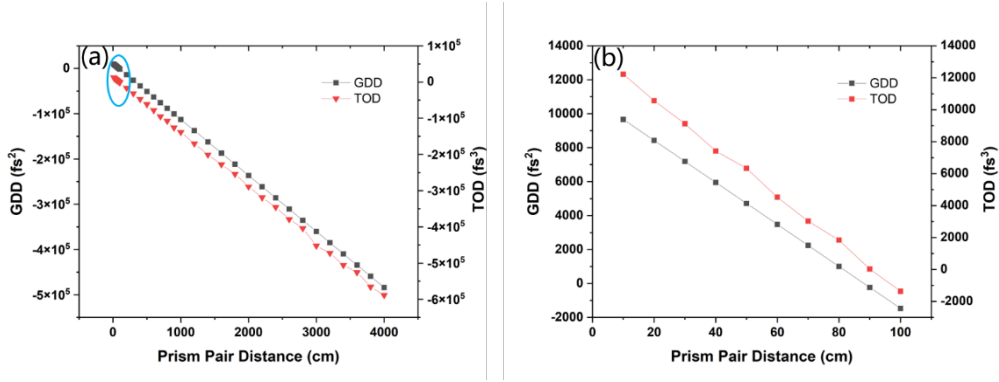

Fig.7 (a) The curves of the second-order and third-order dispersion. (b) is the magnified part marked by the golden blue in (a).

In fact, the loss of the prism pair is low, so its absorption can be ignored. The positive and negative of the second-order dispersion and the third-order dispersion can be precisely adjusted by adjusting the prism insertion. In other words, the total dispersion of a fixed small distance prism pair can be adjusted positively and negatively by changing the material dispersion which can be changed by changing the distance traveled in the prism pair based on formulas (7-9). According to the simulation results, the second-order and third-order dispersion induced by prism pair changes continuously from positive to negative with the increase distance between prism pair. When the prism pair distance reaches the meter-level unit, the second-order and third-order dispersion are both negative. At this time, the positive dispersion introduced by the stretcher part and the material dispersion of the front components will be pre-compensated. Meanwhile, we can fully compensate the negative chirp with thick glass plate[17], so that the laser beam can be shaped in spatial without changing the temporal characteristics. Note that grating pair together with prism pair were always used to full compensate the third-order dispersion of the whole laser system to achieve a short pulse duration with broad spectral bandwidth because the third-order dispersions induced by prism pair and grating pair are opposite [18]. Then, the prism pair added before the gratings based compressor is help to compensate the third-order and even fourth-order spectral dispersion in PW laser [18, 19].

## 5. CONCLUSION

In this study, a method of beam smoothing based on prism pair is proposed, which can reduce the hot spot intensity in high peak power laser beam due to the diffuse reflection of dust or the defects of optical elements. Thereby, using this means can avoid the damage of the first grating and the fourth grating in the grating compression system. The influence factors of smoothing beam based on the prism pair are explored. The results of simulation show that LSIMR is closely related to the induced spatial dispersion width. And the spatial dispersion width has a great relationship with the apex angle of prism pair, the distance between prism pair and the incident of beam.

In this simulation, we not only explore the laser beam smoothing effect of a single prim pair, but also simulate the beam smoothing effect of two prism pairs. The proposed two prism pairs to introduce spatial dispersion can divide into one direction and two directions. The simulation results show that smoothing effect of two prism pairs can achieved bigger spatial dispersion width in a relatively small experiment space. Finally, we simulate the temporal dispersion of the central wavelength after passing through a prism pair. In combine with both grating pair and prism pair will also help to compensate the third-order spectral dispersion to achieve the shortest pulse duration with broad spectral bandwidth.

Prism pair with small apex angle has the advantages of no aberration, simple device, basically no energy loss, and it can be used as image transmission (without beam expansion). Therefore, the use of prism pair smooth beam does not affect the characteristics of laser beam. The method proposed in this study is applicable and of great significance in the field of ultra-short pulse compression system.